\documentstyle[12pt,epsf,rotating]{zer}

\newcommand{\lsim}{\raisebox{-0.13cm}{~\shortstack{$<$ \\[-0.07cm] $\sim$}}~}

\newcommand{\nn}{\noindent}
\newcommand{\non}{\nonumber}
\newcommand{\ctowidth}[2]{ \setbox\mycount=\hbox{$#2$}
                          \hbox to \wd\mycount{$ \hss #1 \hss $} }
\newcommand{\ltowidth}[2]{ \setbox\mycount=\hbox{$#2$}
                          \hbox to \wd\mycount{$\hskip0pt plus0pt minus1fil
                           #1 \hfill $} }
\newcommand{\rtowidth}[2]{ \setbox\mycount=\hbox{$#2$}
                          \hbox to \wd\mycount{$\hfill #1
                          \hskip0pt plus0pt minus1fil$} }
\renewcommand {\rm} {\mathrm}
\newcommand {\ee}         {\mathrm{e}^+\mathrm{e}^-}

\newcommand {\cAt} {\mbox{$\cal A_{\tau}$}}

\newcommand {\alfas}   {\alpha_s}

\begin{document}

\begin{titlepage}

\begin{flushright}
DESY 94--148\\
hep-ph/9409457\\
August 1994 \\
\end{flushright}

\vspace{1.5cm}

\begin{center}

{\Large\sc Bounds on Radii and Magnetic Dipole Moments\\[0.5cm] of
           Quarks and Leptons from LEP, SLC and HERA}

\vspace{2cm}

{\large G.\,K\"opp$^1$, D.\,Schaile$^2$, M.\,Spira$^3$ and P.M.\ Zerwas$^3$}
\vspace{1.5cm}

\nn
{
$^1$ IIIA Phys.~Institut, RWTH, D--52056 Aachen, FRG \\
$^2$ CERN, CH-1211 Geneva 23, Switzerland\\
$^3$ Deutsches Elektronen--Synchrotron DESY, D--22603 Hamburg, FRG
}
\end{center}

\vspace{2cm}

\begin{abstract}
\nn
Leptons, quarks and gauge bosons are assumed to be pointlike particles in
the Standard Model.
Stringent bounds on the radii of quarks and leptons and their
weak anomalous magnetic moments can be derived from the high--precision
measurements at LEP and SLC.
We find a model--independent bound of
$R\lsim 10^{-17}cm$
for quark and lepton radii.
HERA will provide
complementary information on the electromagnetic static properties of
the quarks and the parameters of the charged quark currents.
\end{abstract}

\end{titlepage}

\section{Introduction}
The Standard Model has proven so tremendously sucessful that any
studies of the {\it terra incognita} beyond are highly speculative.
However, the Higgs sector allows conclusions on possible
boundaries of the model. If the Higgs
mass is light, the model can be extrapolated to energy scales of the
order of the Planck mass, yet likely demanding a supersymmetric extension in
the Te$\!$V range.
Also, if
the Higgs mass is heavy, new physical phenomena may emerge at energy
scales in the Te$\!$V range.

The fundamental particles -- leptons and quarks, gauge and Higgs bosons
-- are assumed to be pointlike in the Standard Model.
Possible substructures as well as any new
interactions at high energies would
manifest themselves as non--zero radii and anomalous moments
of these particles.

Deep--inelastic scattering of electrons on protons
is one of the classical tools to probe the static parameters of quarks
\cite{2}. This method was
based originally on the assumption that the photon is elementary and
that electrons
can be treated effectively as pointlike particles down to distances of
order $10^{-17} cm$. Supporting evidence for this assumption follows from the
high--precision measurements of the magnetic dipole
moment of the electron
\cite{3}. If non--standard contributions to the
anomalous magnetic moment, \linebreak
$\delta (g-2)_e  \lsim  10^{-10}$,
scale linearly with the fermion mass, the electron radius is bound to
$R_e \lsim 2 \cdot 10^{-21} cm$.
However, if ${\delta}(g-2)_e$
depends quadratically on the fermion mass
the bound is weakened to
$R_e \lsim 3 \cdot 10^{-16} cm$
and non--pointlike substructures of the electrons have to be
taken into account eventually
when deep--inelastic scattering data are evaluated.
The quadratic mass dependence is a natural
consequence of chiral symmetry \cite{7} which could
keep the fermion masses much smaller than the energy scale
$R^{-1}$ of the substructure and could allow for non-zero axial
couplings for small $R$ \cite{Schlu}.
The radius of the muon is smaller by an order of magnitude
in this scenario.

Complementary information can be extracted from $e^+ e^-$ annihilation into
fermion pairs, in particular on the $Z$ resonance.
While the assumption of elementary photons may be considered
quite natural, the analogous
assumption for the massive $Z$ boson is less obvious; note, however,
that magnetic and quadrupole moments of a composite
spin-one system have been shown \cite{Hill} to
approach the standard-model values in the zero-radius limit.
Possible deviations of the $Z\bar FF$ vertex from the pointlike form may
therefore be attributed to substructures either of the quarks or of the $Z$
boson. However, barring unnatural cancellations, bounds derived
from the data will
apply approximately to both species of particles at the same time.

Setting bounds on anomalous values of the electroweak static parameters
provides an alternative method
to the analysis of contact interactions in $ee$, $eq$
and $qq$ elastic scattering \cite{4}. The bounds on the
energy scales $\Lambda$ of the contact
interactions,
typically about 2 to 5 Te$\!$V for electrons
\cite{5} and above 1 Te$\!$V for quarks \cite{6},
can be transformed into bounds on the radii of electrons and quarks only
after the new strong coupling $g_*$ is fixed, $R \sim \sqrt{4\pi/g_*^2}
\Lambda^{-1}$. $\Lambda$ values of more than 1 Te$\!$V
correspond to radii $R \lsim 6 \times 10^{-17} cm$
within factors of three.

Assuming ${\cal C},{\cal P}$ invariance for the $\gamma$
couplings to fermions and ${\cal CP}$ invariance for the $Z,W$
couplings,
the $\gamma,Z,W$--fermion vertices\footnote[1]{${\cal CP}$ non--invariance
of these interactions gives rise to electric dipole moments which have
recently attracted theoretical \cite{6a}
and experimental attention \cite{6b}.}
$\Gamma_{\lambda}$
are parameterized by form factors, depending on the momentum transfer
$q$, in the following way: \\[-0.2cm]
\begin{minipage}[t]{4.3cm}{
\vspace{1cm}
\hspace*{2.0cm}
\begin{turn}{-90}%
\epsfxsize=7cm \epsfbox{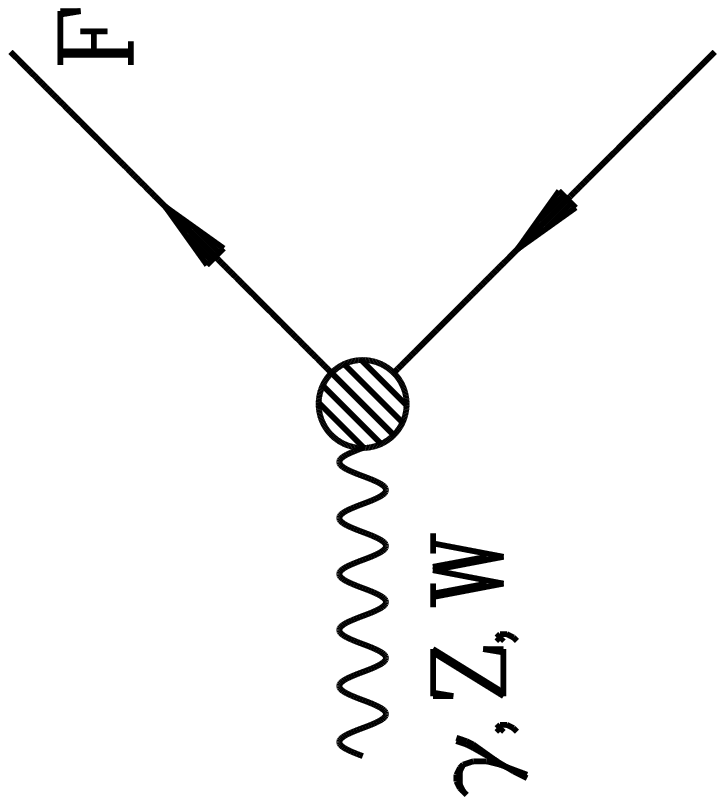}
\end{turn}
\vspace{-4cm}
}
\end{minipage} \hfill
\begin{minipage}[t]{11cm}
\begin{eqnarray}
\Gamma_\lambda^\gamma & = &e_0 e \left[ f\gamma_\lambda +
i \frac{\kappa}{2m_F}
\sigma_{\lambda\rho} q_\rho \right] \non \\
\Gamma_\lambda^Z & = & \sqrt{\frac{G_F m_Z^2}{2\sqrt{2}}}
\left[ v f \gamma_\lambda +
i \frac{\kappa}{2m_F}
v \sigma_{\lambda\rho} q_\rho
- a f \gamma_\lambda \gamma_5 \right] \non \\
\Gamma_\lambda^W & = & \sqrt{\frac{G_F m_W^2}{\sqrt{2}}}\ (1+\gamma_5)
\left[ f \gamma_\lambda +
i \frac{\kappa}{2m_F}
\sigma_{\lambda\rho} q_\rho \right]
\end{eqnarray}
\end{minipage}

\nn
The form factors $f$
reduce to unity and the couplings to
the familiar values in the
Standard Model [$e_0e$; $v=2I_3-4e\sin^2\theta_w$ and $a=2I_3$]
in the pointlike limit.
For the ${\cal CC}$ processes we take the neutrinos in
the final state as purely left--handed particles [and the quarks,
too].
For the sake of simplicity we also
assume the radius $R$ to be
universal for the weak vector and axial--vector probes\footnote[2]{The
form factors $f$ and the anomalous magnetic moments
may depend, in principle, on the vector bosons $\gamma,Z,W$ as well as
on the fermion species.}
so that the form factor is parameterized in the standard way as
\begin{equation}
f = 1 + \textstyle \frac{1}{6} \displaystyle R^2 q^2
\end{equation}
$\kappa$ is the anomalous magnetic dipole moment of the fermion $F$ in units
of the $F$ Bohr magneton $e_0e/2m_F$ {\it etc}. For light $(u,d)$
and $s$ quarks we
identify $m_q$ with the values
$\approx 4$ Me$\!$V and 95 Me$\!$V,
respectively, corresponding to the running current quark
masses defined at the scale
$m_Z$; for charm and bottom quarks we choose 0.7 and 3.0 Ge$\!$V
\cite{6gl}.
For complex systems the magnitude of the anomalous magnetic
dipole couplings
is expected to be of the spatial size of the system \cite{7} so that
$\kappa$ should scale with the $F$ mass.
\begin{table}[hbt]
\begin{center}\begin{tabular}{|c|cc|} \hline
 & linear $F$ mass & quadratic $F$ mass \\
 & dependence      & dependence
\renewcommand{\arraystretch}{1.5}
\\ \hline \hline & & \\[-0.5cm]
$\kappa$ & $\eta_1 m_F R$ & $\left(\eta_2 m_F R\right)^2$ \\
$\displaystyle \frac{\kappa}{2m_F}$ & $\frac{1}{2} \eta_1 R$ &
$\frac{1}{2}  m_F \left(\eta_2 R\right)^2$ \\[0.3cm] \hline
\end{tabular}\end{center}
\renewcommand{\arraystretch}{1}
\caption{\it Definition of the dimensionless parameters
$\eta_1$ and $\eta_2$,
assumed to be independent of the fermion mass.} \label{tb:1}
\end{table}
Since dipole interactions flip
the
chirality of the states, $\kappa$ will scale quadratically with the $F$
mass in chirally symmetric theories \cite{7}.
Particularly in the latter scenario, the heavy fermions $\tau$, $c$, $b$ [and
$t$ in the future \cite{7A}] are of special interest since the anomalous
contributions are enhanced by seven to eight
orders of magnitude over the electron parameters. For the linear and
quadratic $F$ mass scenarios we
therefore define two parameters $\eta_1$ and $\eta_2$, as shown
in Table \ref{tb:1},
which are assumed to be independent of the fermion $F$ mass. The
bounds derived for the electron radius $R_e$
from the measurements of the anomalous
magnetic moment $\kappa_e$ correspond
to the values $\eta_{1,2}=1$ by definition. In high--energy scattering
experiments the particle radius $R$ and the anomalous magnetic dipole moments
can be measured separately so that these two independent parameters can be
disentangled.

\section{$e^+e^-$ Annihilation and $Z$ Decays}
The cross section for the process
$e^- + e^+ \rightarrow F + \bar F$ is mediated by
$s$--channel $\gamma$ and $Z$ boson exchange for $F\neq e$. Neglecting the
fermion masses in a first step [see Ref.\cite{7A,8} for mass corrections],
the cross section can
be written as an incoherent superposition of the helicity cross sections
$\sigma(e^-_i + e^+_j \rightarrow F_k + \bar F_l)$ with $i,\ldots = L,R$:
\begin{equation}
\frac{d\sigma}{d\cos\theta} = \frac{4\pi\alpha^2 N_{\cal C}}{3s}
\frac{1}{4}
\sum_{L,R} \sigma_{ij}^{kl} f_{ij}^{kl} (\cos\theta)
\end{equation}
$s$ denotes the total energy squared and
$\theta$ the angle between the $F$ momentum and
the $e^-$ beam axis; $N_{\cal C}=3$ and 1 for quark and leptons, respectively.
Vector/axial--vector currents [$V$]
are helicity conserving while tensor currents [$\Sigma$]
involving the anomalous magnetic moments, flip the helicity so that
we obtain the following angular distributions
for the annihilation/creation of fermion--antifermion pairs:
\begin{equation}
f_{ij}^{kl} (\cos\theta) = \left\{ \begin{array}{ll}
\frac{3}{8} (1\pm \cos\theta)^2 \hspace{0.5cm} & \mbox{~for~} V_e \times
V_F \mbox{~[hel($F)=\pm$ hel($e^-$)]} \\
\frac{3}{8} \sin^2\theta & \mbox{~for~} V_e \times \Sigma_F
\mbox{~and {\it v.v.}} \\
\frac{3}{8} \cos^2\theta & \mbox{~for~} \Sigma_e \times \Sigma_F
\end{array} \right.
\end{equation}
The form of these coefficients can easily be derived from angular momentum
conservation. Since vector/axial--vector currents couple the particles to
spin--1 states polarized along the flight direction, either backward or
forward production is forbidden [equal and opposite helicities in the
initial and final states]. Since the spin currents $\Sigma$ couple the
particles to spin $S_z =0$ states, the spin of the intermediate
$\gamma,Z$
bosons points into the direction perpendicular to the fermion
flight axes
so that $F$ production at $90^0$ cannot occur in this case. On the other
hand, angular momentum conservation forbids non--zero values of the
$V \times \Sigma$ interference terms for forward and backward production
since the spin $S_z$ would change along the beam axis by one unit
without being balanced by orbital angular momenta.

Introducing the generalized charges, related to vector/axial--vector and spin
currents,
\begin{equation}
\begin{array}{llll}
Q_i^k & = &
\displaystyle
e_e e_F + \frac{G_F m_Z^2}{8\sqrt{2}\pi\alpha} \epsilon_e^i
\epsilon_F^k \frac{s}{s-m_Z^2+im_Z\Gamma_Z} \hspace{0.5cm} & \mbox{~for~}
V_e \times V_F \\ \\
Q'_i & = &
\displaystyle
e_e e_F + \frac{G_F m_Z^2}{8\sqrt{2}\pi\alpha} \epsilon_e^i
v_F \frac{s}{s-m_Z^2+im_Z\Gamma_Z} & \mbox{~for~} V_e \times \Sigma_F \\ \\
Q'^k & = &
\displaystyle
e_e e_F + \frac{G_F m_Z^2}{8\sqrt{2}\pi\alpha} v_e \epsilon_F^k
\frac{s}{s-m_Z^2+im_Z\Gamma_Z} & \mbox{~for~} \Sigma_e \times V_F \\ \\
Q'' & = &
\displaystyle
e_e e_F + \frac{G_F m_Z^2}{8\sqrt{2}\pi\alpha} v_e
v_F \frac{s}{s-m_Z^2+im_Z\Gamma_Z} & \mbox{~for~} \Sigma_e \times \Sigma_F
\end{array}
\end{equation}
with $\epsilon^{L,R} = v \pm a$, the coefficients $\sigma_{ij}^{kl}$ are
given by
\begin{equation}
\sigma_{ij}^{kl} = \left\{ \begin{array}{ll}
|Q_i^k|^2 f_e^2 f_F^2 & \mbox{~for~} V_e \times V_F \\
|Q'_i|^2 f_e^2 \frac{1}{4} (\kappa_F/m_F)^2 s & \mbox{~for~} V_e \times
\Sigma_F \\
|Q'^k|^2 \frac{1}{4} (\kappa_e/m_e)^2 s f_F^2 & \mbox{~for~} \Sigma_e
\times V_F \\
|Q''|^2 \frac{1}{16} (\kappa_e/m_e)^2 (\kappa_F/m_F)^2 s^2
\hspace{0.5cm} & \mbox{~for~} \Sigma_e \times \Sigma_F
\end{array} \right.
\end{equation}

Non--zero radii and anomalous magnetic moments affect the total
cross sections and the angular distributions of the produced
fermions.\footnote{Note that $\kappa_F/m_F\sim R$ or $m_F R^2$ does
not rise with the inverse fermion mass so that the
limit $m_F \to 0$ can safely be applied everywhere, in particular for
electrons.}

On top of the $Z$ boson, three observables are of particular interest --
the
partial width $\Gamma(Z\rightarrow F\bar F)$, the forward--backward
asymmetry of the leptons/quark jets and the $\alpha_F$ parameter,
defined by $dN/d\cos\theta
\sim 1+\alpha_F \cos^2\theta + \beta_F \cos \theta$ and measuring the
strength of the longitudinal
cross section. Deviations from the [improved] Born cross sections,
marked by the index $B$, may be
expressed in terms of the electron and quark/lepton
radii $R_F$ and the anomalous moments
$\kappa_F$ in the following way [$F$ mass terms included]:
\begin{eqnarray}
\Gamma(Z\rightarrow F\bar F) & = & \Gamma_B(Z\rightarrow F\bar F) \left[
1+
\textstyle
\frac{1}{3}
\displaystyle
(m_Z R_F)^2 + 3\kappa_1 \right] \non \\ \non \\
A_{FB}(F) & = & A_{FB}^B \left[ 1+ \kappa_2 \right] \non \\ \non \\
\alpha_F & = & 1-4 \kappa_3
\end{eqnarray}
Additional information is provided by the left--right asymmetry for polarized
electrons/po\-si\-trons, as well as the LR asymmetry of the polarized $\tau$'s
in the final state.
\begin{eqnarray}
{\cal A}_e & = & {\cal A}_e^B \left[ 1+ \kappa_4 \right] \non \\ \non \\
{\cal A}_\tau & = & {\cal A}_\tau^B \left[ 1+
\textstyle
\frac{3}{2}
\displaystyle
\kappa_5 \right]
\end{eqnarray}
The coefficients $\kappa_i$ are given by
\begin{eqnarray}
\kappa_1 & = & \kappa_F
\left\{ 1 +
\textstyle
\frac{1}{24}
\displaystyle
\kappa_F \left({m_Z}/{m_F}\right)^2
\right\} \hat v_q^2 \non \\ \non \\
\kappa_2 & = & \kappa_F \left\{ 1-3\hat v_F^2 -
\textstyle
\frac{1}{8}
\displaystyle
\kappa_F \left({m_Z}/{m_F}\right)^2 \hat v_F^2 \right\} -
\textstyle
\frac{1}{8}
\displaystyle
\kappa_e^2 \left({m_Z}/{m_e}\right)^2 \hat v_e^2
\non \\ \non \\
\kappa_3 & = & \kappa_F \left\{ 1 +
\textstyle
\frac{1}{8}
\displaystyle
\kappa_F \left({m_Z}/{m_F}\right)^2 \right\} \hat v_F^2 +
\textstyle
\frac{1}{8}
\displaystyle
\kappa_e^2 \left({m_Z}/{m_e}\right)^2 \hat v_e^2
\end{eqnarray}
and, for the LR asymmetries,
\begin{eqnarray}
\kappa_4 & = & -
\textstyle
\frac{1}{8}
\displaystyle
\kappa_e^2 \left({m_Z}/{m_e}\right)^2
\hat v_e^2 \non \\ \non \\
\kappa_5 & = & \kappa_\tau \left\{ 1-4\hat v_\tau^2 -
\textstyle
\frac{1}{12}
\displaystyle
\kappa_\tau \left({m_Z}/{m_\tau}\right)^2 \hat v_\tau^2 \right\}
\end{eqnarray}
with $\hat v_F^2 = v_F^2/(v_F^2 + a_F^2)$ {\it etc}.

To give a flavour of the sensitivity of $\ee$ collider data at the
$Z$ we base our evaluation on the averages of published data as
summarized in~\cite{bib-PDG}.
We use the
following measurements as input to our fits:
The mass of the $Z$, $M_Z=91.187\pm0.007$~GeV, the total width of
the $Z$, $\Gamma_Z=2.490\pm0.007$~GeV,
the hadronic pole cross section, $\sigma_h^0=41.55\pm0.14$~nb,
the ratio of the hadronic
partial width to the partial width for $Z$ decays into electron,
muon and tau pairs, $R_e=20.76\pm0.08$, $R_\mu=20.76\pm0.07$ and
$R_\tau=20.80\pm0.08$, the forward--backward
pole asymmetries for electrons, muons and taus,
$A_{FB}^{0,e}=0.0151\pm0.040$,
$A_{FB}^{0,\mu}=0.0133\pm0.0026$ and
$A_{FB}^{0,\tau}=0.0212\pm0.0032$, the integrated tau
polarization asymmetry, $\cAt=0.141\pm0.021$, the combination
of the tau polarization
forward--backward asymmetry and the measurement of the left-right
polarization asymmetry at SLC, ${\cal A}_e=0.161\pm0.012$,
the forward--backward pole asymmetries for $b$ and $c$ quarks,
$A_{FB}^{0,b}=0.107\pm0.013$ and $A_{FB}^{0,c}=0.058\pm0.022$
and the ratio of the
partial width of the $Z$
for decays into $b$ quarks to the hadronic partial width,
$\Gamma_{b\bar b}/\Gamma_{had}=0.2210\pm0.0029$.

For the sake of simplicity we assume the radius $R$ to be
independent of the particle species. For the evaluation of the
Standard Model prediction we use the electroweak library provided
by the program ZFITTER~\cite{bib-ZFITTER}. The mass of the
top quark $m_t$ is treated as an additional free parameter.
Also the strong coupling constant $\alfas$ is treated as free parameter
within the limits $\alpha(m_Z^2)
=0.123\pm0.006$~\cite{bib-LEPQCD}.
As the results
presented below are not sensitive to a variation of the mass
of the Higgs boson between 60 and 1000~Ge$\!$V, this unknown
parameter has been fixed to $m_H=300$~Ge$\!$V.

We find the values listed in Table~\ref{tb:2}
for the anomalous magnetic moments in the
linear and quadratic $F$ mass scenarios.
In the bottom part of the table we
compare the results with the corresponding bounds derived from the
$(g-2)$
measurements~\cite{3}. The $Z$ decay data are all compatible
with a vanishing fermion radius and with no anomalous contributions
to the magnetic moments.
\begin{table}[hbt]
\begin{center}
\begin{tabular}{|c||c|c||c|c|} \hline
 & \multicolumn {2} {c||} {linear $F$ mass}
 & \multicolumn {2} {c|}  {quadratic $F$ mass} \\
 & \multicolumn {2} {c||} {dependence of $\kappa$}
 & \multicolumn {2} {c|}  {dependence of $\kappa$} \\ \hline \hline
$ZF\overline{F}$
 & \multicolumn {2} {c||} {$R=(0.0\pm 0.1)\cdot 10^{-3}~fm$}
 & \multicolumn {2} {c|}  {$R=(0.0\pm 0.1)\cdot 10^{-3}~fm$}\\ \hline
 & $\eta_1R~[fm]$ & $\kappa$ &
   $\eta_2R~[fm]$ & $\kappa$ \\ \hline
$\rm e$ &
$(-0.4_{-0.2}^{+1.1})\cdot 10^{-2}$&
$(-0.2_{-0.1}^{+0.5})\cdot 10^{-5}$&
$1.3_{-2.8}^{+0.3}$&
$(0.4_{-1.9}^{+0.2})\cdot 10^{-6}$ \\
$\mu$ &
$(-0.4_{-0.3}^{+1.0})\cdot 10^{-2}$&
$(-0.4_{-0.3}^{+1.1})\cdot 10^{-3}$&
$(0.8_{-1.9}^{+0.3})\cdot 10^{-1}$ &
$(0.7_{-3.4}^{+0.6})\cdot 10^{-4}$\\
$\tau$ &
$(0.4_{-0.7}^{+0.3})\cdot 10^{-2}$&
$(0.7_{-1.2}^{+0.5})\cdot 10^{-2}$&
$(0.2_{-0.5}^{+0.1})\cdot 10^{-1}$&
$(0.1_{-0.6}^{+0.1})\cdot 10^{-2}$\\
$u,d$ &
$(0.0\pm0.6)\cdot 10^{-3}$&
$(0.0\pm0.2)\cdot 10^{-5}$&
$(0.0\pm0.4)\cdot 10^{-1}$&
$(0.0\pm0.3)\cdot 10^{-10}$\\
$s$ &
$(0.0\pm0.6)\cdot 10^{-3}$&
$(0.0\pm0.5)\cdot 10^{-4}$&
$(0.0\pm0.4)\cdot 10^{-1}$&
$(0.0\pm0.2)\cdot 10^{-7}$\\
$c$ &
$(0.0\pm0.2)\cdot 10^{-2}$&
$(0.0\pm0.1)\cdot 10^{-2}$&
$(0.0\pm0.2)\cdot 10^{-1}$&
$(0.0\pm0.4)\cdot 10^{-6}$ \\
$b$ &
$(0.6\pm0.5)\cdot 10^{-3}$&
$(0.2\pm0.2)\cdot 10^{-2}$&
$(0.6_{-0.5}^{+0.2})\cdot 10^{-2}$&
$(0.4_{-0.6}^{+0.2})\cdot 10^{-3}$  \\ \hline \hline
$\gamma {\rm e} \bar {\rm e}$
& \multicolumn {4} {c|}  {$\kappa = (0.5\pm0.3)\cdot 10^{-10}$}\\
$\gamma {\mu} \bar {\mu}$
& \multicolumn {4} {c|}  {$\kappa = (0.8\pm1.1)\cdot 10^{-8}$}\\ \hline
\end{tabular}
\end{center}
\caption[foo]{ \label{tb:2} \it
Bounds on the particle radius $R$ and
the anomalous magnetic moments $\kappa$. Upper part:
bounds derived from LEP $Z$ decays [$\chi^2/d.o.f. = 10/5$].
Lower part: bounds from $(g-2)$ measurements; they are defined by the
difference between the theoretical and experimental average values with
their errors added in quadrature.
}
\end{table}

The analysis of the heavy $\tau$ final states is of special interest
in this context \citer{8,9a}. Early PETRA data have been used \cite{9}
to constrain the anomalous magnetic dipole moment of the $\tau$ to less
than $0.014$. A bound of $0.036$ may finally be obtained from $\gamma$
radiation in $\tau$ pair production of $Z$ decays \cite{9a}. It is
apparent from
Table \ref{tb:2} that the bound on the $Z$ anomalous dipole moment is stronger
than these estimates.

As expected, also the bounds on the anomalous magnetic moments of the
light quarks improve considerably compared to earlier evaluations of
PETRA data \cite{9,10}. The limits on the anomalous magnetic
moments of the heavy $c,b$ quarks are much stronger than the bounds
obtained from the low--energy data.

\section{Electron--Quark Scattering}
\subsection{${\cal NC}$ Processes}
The cross sections for electron--quark scattering
$e^\pm \mbox{\shortstack{$\scriptstyle (-)$ \\[-0.1cm]
$q$}} \rightarrow e^\pm \mbox{\shortstack{$\scriptstyle (-)$ \\[-0.1cm] $q$}}$
at HERA energies are built--up by
photon and $Z$--boson exchanges.  For large values of the
momentum transfer, $Q^2$ of order $10^4$Ge$\!$V$^2$,
the $Z$--exchange contributions are of the same order as the
$\gamma$ contributions. Restricting ourselves to light quark targets,
the cross section $ep\rightarrow eX$ can be decomposed into the
following incoherent sum of helicity cross sections,
\begin{equation}
\frac{d\sigma^{\cal NC}}{dx dQ^2} = \frac{4\pi\alpha^2}{Q^4} \frac{1}{4} \sum_q
\sum_{L,R} q(x,Q^2) \sigma_{ij}^{kl} f_{ij}^{kl} (y)
\end{equation}
(including quark and antiquark parton densities $q$).
$y$ denotes the usual energy transfer variable $pq/pk_e$.
Defining the
generalized charges for vector/axial--vector and spin currents
in the process $e^-_i + q_k \rightarrow e^-_j + q_l$ as
\begin{equation}
\begin{array}{llll}
Q_i^k & = &
\displaystyle
e_e e_q + \frac{G_F m_Z^2}{8\sqrt{2}\pi\alpha} \epsilon_e^i
\epsilon_q^k \frac{Q^2}{Q^2+m_Z^2} \hspace{0.5cm} & \mbox{~for~} V_e \times
V_q \\ \\
Q'_i & = &
\displaystyle
e_e e_q + \frac{G_F m_Z^2}{8\sqrt{2}\pi\alpha} \epsilon_e^i
v_q \frac{Q^2}{Q^2+m_Z^2} & \mbox{~for~} V_e \times \Sigma_q \\ \\
Q'^k & = &
\displaystyle
e_e e_q + \frac{G_F m_Z^2}{8\sqrt{2}\pi\alpha} v_e \epsilon_q^k
\frac{Q^2}{Q^2+m_Z^2} & \mbox{~for~} \Sigma_e \times V_q \\ \\
Q'' & = &
\displaystyle
e_e e_q + \frac{G_F m_Z^2}{8\sqrt{2}\pi\alpha} v_e
v_q \frac{Q^2}{Q^2+m_Z^2} & \mbox{~for~} \Sigma_e \times \Sigma_q
\end{array}
\end{equation}
the coefficients $\sigma_{ij}^{kl}$ are given by
\begin{equation}
\sigma_{ij}^{kl} = \left\{ \begin{array}{ll}
|Q_i^k|^2 f_e^2 f_q^2 & \mbox{~for~} V_e \times V_q \\
|Q'_i|^2 f_e^2 \frac{1}{4} (\kappa_q/m_q)^2 s_{eq} & \mbox{~for~}
V_e \times \Sigma_q \\
|Q'^k|^2 \frac{1}{4} (\kappa_e/m_e)^2 s_{eq} f_q^2 & \mbox{~for~}
\Sigma_e \times V_q \\
|Q''|^2 \frac{1}{16} (\kappa_e/m_e)^2 (\kappa_q/m_q)^2 s^2_{eq}
\hspace{0.5cm} & \mbox{~for~} \Sigma_e \times \Sigma_q
\end{array} \right.
\end{equation}
where $s_{eq}$ denotes the invariant energy squared of the $eq$
subprocess.
The $y$ dependent coefficients are related to the scattering angle,
$y=(1+\cos\theta_*)/2$, in the $(eq)$ c.m.~frame,
\begin{equation}
f_{ij}^{kl} (y) = \left\{ \begin{array}{ll}
\displaystyle
1 & \mbox{~for~} V_e \times V_q \mbox{~and equal $e/q$ helicities} \\
(1-y)^2 & \mbox{~for~} V_e \times V_q \mbox{~and opposite $e/q$ helicities}
\\
\displaystyle
y(1-y)  & \mbox{~for~} V_e \times \Sigma_q \mbox{~and {\it v.v.}} \\
y^2 (1-\frac{1}{2}y)^2 \hspace{0.5cm} & \mbox{~for~} \Sigma_e \times \Sigma_q
\end{array} \right. \label{eq:13}
\end{equation}
Switching from electrons to positrons or/and quarks to antiquarks, the helicity
factors $f$ remain the same while the generalized charges of the
antiparticles are to
be identified with the negative charges of the particles, yet the
helicity indices reversed.
The first two entries in eq.(\ref{eq:13}) are well--known consequences
of angular momentum
conservation for helicity--conserving vector/axial--vector couplings.
Since the spin couplings flip the helicities, the interference term between
$V$ and $\Sigma$ couplings must vanish for forward and backward scattering;
this condition is met by the coefficient $y(1-y)=(1-\cos^2\theta_*)/4$.
Fierzing the $\Sigma \times \Sigma$ amplitude from $t$ to $s$--channel
exchange amplitudes
results in a mixture of spin--0 and spin--1 amplitudes so that no
simple angular pattern can be derived in this case.

Note that the contributions due to the anomalous magnetic moments come
with characteristic $y(1-y)$ and $y^2(1-\frac{1}{2}y)^2$ coefficients
which are
different from the familiar [$1\pm (1-y)^2$] coefficients of the leading $V,A$
currents and the $y^2$ coefficient of the non--leading longitudinal
structure function. Non--zero quark radii, on the other hand, affect the
$Q^2$ dependence of the deep inelastic cross section in a characteristic way.

Assuming that the
cross section $d\sigma (eP \rightarrow eX)/dQ^2$ can be measured at HERA with
an accuracy of order 1\% for $Q^2$ values of order $10^3$ Ge$\!$V$^2$, the
radius of the light quarks can be limited to
$R_q \lsim 10^{-16} cm$.
Note that the $\gamma$ exchange dominates the cross section in this $Q^2$
range. This bound is therefore complementary to the bound extracted from
$Z$ decay data, and can truly be associated with the quark radius.
If the
anomalous $y$ dependent terms can also be limited to a level of 1\%, the
$\kappa_q$ parameter for quarks is restricted to less than
$\kappa_q \lsim 10^{-4}$
for light quarks $u,d$.

\subsection{${\cal CC}$ Processes}
The cross sections for the charged current processes
$e^\pm \mbox{\shortstack{$\scriptstyle (-)$ \\[-0.1cm]
$q$}} \rightarrow \mbox{\shortstack{$\scriptstyle (-)$ \\[-0.1cm]
$\nu$}} \mbox{\shortstack{$\scriptstyle (-)$ \\[-0.1cm] $q$}$'$}$
can be written in a similar compact form,
\begin{equation}
\frac{d\sigma^{\cal CC}}{dx dQ^2} = \frac{G_F^2}{2 \pi} \left[
\frac{m_W^2}{Q^2+m_W^2} \right]^2
\sum_q \sum_{L,R} q(x,Q^2) \sigma_{iL}^{kL} f_{iL}^{kL} (y)
\end{equation}
If only the left--handed neutrinos and final state quarks contribute to
the coefficients $\sigma_{ij}^{kl}$ [$j=l=L$] of $\sigma(
e_i^\pm q_k \rightarrow \nu_j q'_l)$, we obtain
\begin{equation}
\sigma_{iL}^{kL} = \left\{ \begin{array}{ll}
f_e^2 f_q^2 & \mbox{~for~} V_e \times V_q \\
f_e^2 \frac{1}{4} (\kappa_q/m_q)^2 s_{eq} & \mbox{~for~}
V_e \times \Sigma_q \\
\frac{1}{4} (\kappa_e/m_e)^2 s_{eq} f_q^2 & \mbox{~for~}
\Sigma_e \times V_q \\
\frac{1}{16} (\kappa_e/m_e)^2 (\kappa_q/m_q)^2 s^2_{eq}
\hspace{0.5cm} & \mbox{~for~} \Sigma_e \times \Sigma_q
\end{array} \right.
\end{equation}
The $y$ dependent coefficients agree with the expressions in eq.(\ref{eq:13}).

First measurements of the ${\cal CC}$ cross section at HERA have been
carried out recently \cite{17}.
Bounds on the radii and the
transition magnetic moments of the quarks from ${\cal CC}$ processes
are expected in the same range as for ${\cal NC}$ processes when
the statistics at HERA will grow in the years to come.
 \newpage

\section{Summa}
\begin{table}[hbt]
\begin{center}\begin{tabular}{|c||cc|} \hline
$R~[fm]$ & linear mass & quadratic mass \\
         & dependence of $\kappa$ & dependence of $\kappa$
\renewcommand{\arraystretch}{1.5}
\\ \hline \hline & & \\[-0.5cm]
$Z F \overline{F}$ &
$(0.0\pm0.1)\cdot 10^{-3}$&
$(0.0\pm0.1)\cdot 10^{-3}$ \\[0.1cm] \hline
$\gamma e \bar e$
&$(0.2\pm0.1)\cdot 10^{-7}$
&$(0.3\pm0.1)\cdot 10^{-2}$ \\
$\gamma \mu \bar \mu$
&$(0.1\pm0.2) \cdot 10^{-7}$
&$(0.2\pm0.1)\cdot 10^{-3}$\\[0.1cm] \hline
\end{tabular}\end{center}
\renewcommand{\arraystretch}{1}
\caption{\it Bounds on the radii of the particles in the
Standard Model, derived from the LEP $Z$ experiments and the $(g-2)_{e,\mu}$
measurements. Bounds on anomalous magnetic dipole moments are converted into
bounds on radii $R$ by setting $\kappa_F=m_F R$ and $(m_F R)^2$ for the first
and second column, respectively. For the fit to the $Z$ data
$\chi^2/d.o.f.=12/11$.} \label{tb:3}
\end{table}
\nn
We may summarize the results presented in the previous sections by
setting $\eta_1$ and $\eta_2 = 1$ and converting the bounds on the
anomalous dipole moments into bounds on the radius $R$ of the fermions.
The results on $R$ are collected in Table \ref{tb:3} for the high energy
$Z$ decay experiments. They are compared with the values of $R$
for electrons and muons  derived from $(g-2)$ experiments
in the same way. In the
quadratic $F$ mass scenario the bounds on the radii of leptons and
quarks from $Z$ decays are competitive with the bounds derived from
the $(g-2)_\mu$ experiments.
In this scenario fermions appear to have radii
\begin{equation}
R\lsim 10^{-17}cm
\end{equation}
HERA is expected to set bounds of similar size on quarks and
electrons in
the electromagnetic and weak sectors separately.

\vspace{1.5cm}
\nn {\bf Acknowledgement.} We are grateful to S.J.Brodsky for valuable
discussions and comments on the manuscript.

\end{document}